\documentclass[preprint,authoryear,5p,times,twocolumn]{elsarticle}
\usepackage{graphicx}
\usepackage{amssymb}
\usepackage{amsmath}

\journal{New Astronomy Reviews}

\begin{document}

\begin{frontmatter}

\title{Influence of gravitational microlensing on broad absorption lines of QSOs:
The case of the Fe K$\alpha$ line}

\author[aob]{Predrag Jovanovi\'{c}}
\ead{pjovanovic@aob.bg.ac.rs}
\author[aob]{Luka \v{C}. Popovi\'{c}}
\ead{lpopovic@aob.bg.ac.rs}
\author[fsk]{Sa\v{s}a Simi\'{c}}
\ead{ssimic@kg.ac.rs}
\address[aob]{Astronomical Observatory, Volgina 7, 11060 Belgrade, Serbia}
\address[fsk]{Faculty of Science, University of Kragujevac, Radoja Domanovi\'{c}a 12, 34000 Kragujevac, Serbia}

\begin{abstract}

Here we give a brief overview of some investigations of the
gravitational microlensing influence on broad absorption spectral
lines of lensed QSOs. Especially, we consider the microlensing
influence on the Fe K$\alpha$ broad absorption lines using a model
of an accretion disk covered by an absorption region. Gravitational
microlensing is modeled by ray shooting method which enables us to
obtain realistic microlensing patterns. We obtain that microlensing
can affect both emission and absorption component of line that
depends on dimensions on emission and absorption line regions. Here
we give detailed analysis of emission and absorption line shape
variations due to gravitational microlensing.

\end{abstract}


\end{frontmatter}

\section{Introduction}

Around 10\% of type I (broad emission line) Active Galactic Nuclei
(AGN) show gas outflowing from their centers \citep{rich03} with
velocities $\sim$ 10$^{3-4}$ km s$^{-1}$ (obtained from blue-shifted
absorption lines). Such outflows are probably launched from the
central engine of quasars \citep{el00}. The broad absorption lines
(BALs) can be detected mainly in the UV part of spectra, but also in
Fe K$\alpha$ line \citep{chart04,chart07,done07}.

Gravitational microlensing has proven to be a powerful probe of the
structure at the heart of quasars and therefore microlensing of
Broad Line Absorption Regions (BALR) can be used to study the BALR
structure \citep[see e.g.][etc.]{lew98,ch03,lew03,ch05}.

The X-ray emission of AGN could be significantly absorbed by an
outflowing wind, especially in the case of so-called Low Ionization
Broad Absorption Line (LoBAL) quasars. Presence of such X-ray
absorbers is confirmed in gravitationally lensed LoBAL quasar
H1413+117 (Cloverleaf) at z = 2.56 \citep{chart07}. \citet{chart07}
found that spectra of this quasar show emission features redward of
the Fe K$\alpha$ line rest-frame energy and broad absorption
features blueward of this energy. The line is only significant in
the brighter image $A$ and a microlensing event could explain its
energy-dependent magnification. A cluster of galaxies at redshift $z
= 1.7$ contributes to the lensing of this system, but the absorption
from the lens does not have any significant effect on spectral
properties of H1413+117 while, at the same time, its intrinsic
absorption is much larger and it could significantly affect the
X-ray emission of this quasar \citep{chart04}.

\begin{figure*}
\centering
\includegraphics[angle=-90,width=16cm]{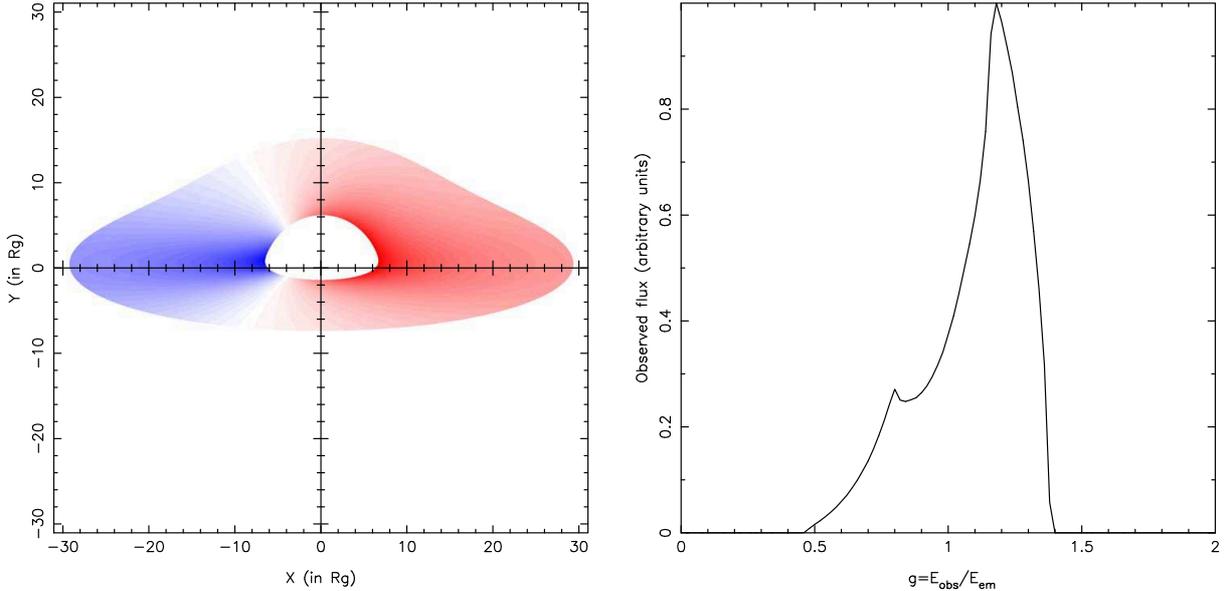}
\caption{The modeled accretion disk (left) and the corresponding Fe
K$\alpha$ emission (right).} \label{fig1}
\end{figure*}

In several previous papers
\citep{lc01,pop03,lcp06,pop06,pj07,pj08,pj09} we have shown that the
effect of strong gravitational lensing can have influence on the
intensity and shape of Fe K$\alpha$ line. Consequently, this effect
can be used as a probe for the structure of the innermost part of
lensed quasars.

The focus of this work  is to study how much the combined influence
of intrinsic absorption and gravitational microlensing could change
the Fe K$\alpha$ spectral line profile, emitted from X-ray emitting
region of LoBAL QSOs.  The paper is organized as following: in \S 2
we describe how to model the X-ray emission of accretion disk of
QSOs, in \S 3 and 4 the combined influence of intrinsic absorption
and gravitational microlensing on this emission is given, in \S 5 we
describe the adopted parameters of X-ray emitting and absorption
regions, and also those which are used for generating the
microlensing pattern for image $A$ of H1413+117. In \S 6 we present
the main results of our numerical simulations and finally in \S 7 we
outline our conclusions.

\section{Modeling the X-ray emission of accretion disk}

The X-ray radiation of AGN is originating from the innermost part of
the accretion disk and the shape of the most prominent spectral line
in this range, Fe K$\alpha$, strongly depends on emissivity law of
the disk. It is usually accepted that surface emissivity of the disk
varies with radius as a power law: $\varepsilon (r) = \varepsilon _0
\cdot r^q,$ where $\varepsilon _0$ is emissivity constant and $q$ is
emissivity index. We modeled the emission of an optically thick and
geometrically thin accretion disk around a supermassive black hole
(SMBH) using numerical simulations based on a ray-tracing method in
Kerr metric, taking into account only photon trajectories reaching
the observer's sky plane in the infinity. In this method one divides
the image of the disk on the observer's sky into a number of small
elements (pixels). For each pixel, the photon trajectory is traced
backward from the observer by following the geodesics in Kerr
space-time, until it crosses the plane of the disk. Then, flux
density of radiation emitted by the disk at that point, as well as
redshift factor of the photon are calculated. In that way, one can
obtain color images of the accretion disk (see Fig. \ref{fig1},
left) which a distant observer would see by a high resolution
telescope. The simulated line profiles can be calculated taking into
account the intensities and received photon energies of all pixels
of the corresponding disk image (see Fig. \ref{fig1}, right). For
more details about this method see e.g. \citet{pj09} and references
therein.

\begin{figure*}
\centering
\includegraphics[width=16cm]{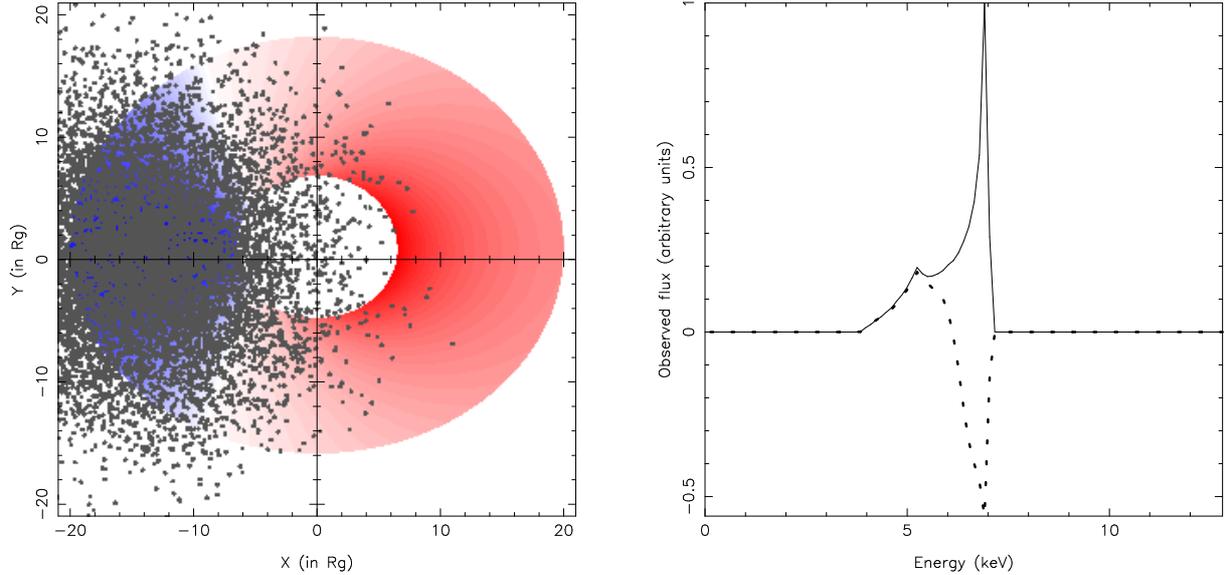}
\caption{The modeled accretion disk covered by warm absorbing matter
(left) and the corresponding Fe K$\alpha$ absorption, presented by
dashed line (right); the solid line on right panel presents the Fe
K$\alpha$ without absorption.}
\label{fig2}
\end{figure*}

\section{A simple model of Fe K$\alpha$ absorption region}

There are different models of the X-ray absorbing/obscuring regions,
like absorbing medium comprised of cold absorbing cloudlets by
\citet{fuerst04}, but here we will focus on the model given by
\citet{pj07}. In this model, absorption region is considered to be
composed of a number of individual spherical absorbing clouds with
the same small radii, scattered in space so that projections of
their centers to the observer's sky plane $(X_i, Y_i)$ have
bivariate normal distribution ${\cal N}_2\left(\mu,\Sigma\right)$.
Here,
$\mu=\left[\mu_X,\mu_Y\right]^T$ and
$$\Sigma=\left[
{\begin{array}{*{20}c}
   {\sigma _X^2 } & {\rho \sigma _X \sigma _Y }  \\
   {\rho \sigma _X \sigma _Y } & {\sigma _Y^2 }  \\
\end{array}} \right]$$
 where $\mu_X$ and $\mu_Y$ are the means of $X_i$ and
$Y_i$, $\sigma _X$ and $\sigma _Y$ are their standard deviations and
$\rho$ is the correlation between them. We use the following values
of these parameters in our numerical simulations: $\rho=0$,
$\mu_X=X_A$, $\mu_Y=Y_A$ and $\sigma_X = \sigma_Y = R_A$, where
$(X_A, Y_A)$ is the center and $R_A$ is the radius of projection of
entire absorption region.

The absorption coefficient $A(X,Y)$ for every spherical cloud in the
absorption region is given by \citep{pj07}:
\begin{equation}
A(X,Y) = \left( 1-I_A(X,Y)\right) \cdot e^{-\left(\dfrac{g(X,Y)\
E_0-E_A}{\sigma_E}\right)^2},
\end{equation}
where absorption intensity coefficient $I_A(X,Y)$ describes the
distribution of absorption over the whole region, $E_A$ is the
central energy of absorption and $\sigma_E$ is the width of
absorption band (velocity dispersion).

To compare (qualitatively) our simulated profiles (Fig. \ref{fig2},
right) with observations, let us recall the results obtained by
\citet{done07}. They found an evidence for so-called P Cygni profile
of the Fe K$\alpha$ line in narrow line Seyfert 1 galaxies.
According to these authors, complex X-ray spectra of these objects
show strong "soft excess" below 2 keV and a sharp drop at $\sim 7$
keV which can be explained either by reflection or by absorption
from relativistic, partially ionized material close to the black
hole. They showed that a sharp feature at $\sim 7$ keV results from
absorption/scattering/emission of iron K$\alpha$ line in the wind.
In the case of 1H 0707-495, this absorption feature can be
satisfactorily fitted by the P Cygni profile, where emission
component at $\sim 5$ keV is followed by a strong absorption
component at $\sim 7$ keV \citep{done07}.

In our model, we are able to change density of absorbers, size of
absorption region (or covering fraction of disk emission) and also
position of absorption region in respect to disk center. As an
example, in Fig. \ref{fig2} (right) we show a deformation of the Fe
K$\alpha$ line profile due to absorption region that covers
approaching side of the disk (Fig. \ref{fig2} left). As one can see
from Fig. \ref{fig2}, in that case a P Cygni profile of the Fe
K$\alpha$ is present, having an emission component at 5 keV and
absorption one at 7 keV, as it was reported by \citet{done07}.

\section{Gravitational microlensing by a random star field in the
lens plane}

The most realistic approximation for gravitational microlensing
amplification is so-called quadrupole microlens. This model is
applied to obtain a spatial distribution of magnifications in the
source plane (where an accretion disk of AGN is located), produced
by a random star field placed in the lens plane \citep[see
e.g.][]{wam92}. Such spatial distribution of magnifications is
called microlensing map, microlensing pattern or caustic network. If
we consider a set of $N$ compact objects (e.g. stars) which are
characterized by their positions $x_i$ and their masses $m_i$, then
normalized lens equation is given by \citep{pj09}:
\begin{equation}
\vec y = \sum\limits_{i = 1}^N {m_i \frac{{\vec x - \vec x_i
}}{{\left| {\vec x - \vec x_i } \right|^2 }}} + \left[
{\begin{array}{*{20}c} {1 - \kappa_c  + \gamma } & 0  \\ 0 & {1 -
\kappa_c  - \gamma }  \\ \end{array}} \right]\vec x,
\end{equation}
where $\vec x$ and $\vec y$ are normalized image and source
positions, respectively. The sum describes light deflection by the
stars and the last term is a quadrupole contribution from galaxy
containing the stars, where $\kappa_c$ is a smooth surface mass
density and $\gamma$ is an external shear. The total surface mass
density or convergence can be written as
$\kappa=\kappa_\ast+\kappa_c$, where $\kappa_\ast$ represents the
contribution from the compact microlenses. The corresponding
microlensing map is then defined by two parameters: the convergence
- $\kappa$, and the shear due to the external mass - $\gamma$. For
some specific microlensing event one can model the corresponding
magnification map using numerical simulations based on ray-shooting
techniques \citep[see e.g.][and references
therein]{pop06,pj08,pj09}, in which the rays are shot from the
observer to the source, through a randomly generated star field in
the lens plane. These light rays are then collected in pixels in the
source plane, and the number of rays in one pixel is proportional to
the magnification due to microlensing at this point in the source
plane. As an example, the magnification map of Cloverleaf quasar
image $A$ is shown in Fig. \ref{fig3}.

\begin{figure}[ht!]
\centering
\includegraphics[width=\columnwidth]{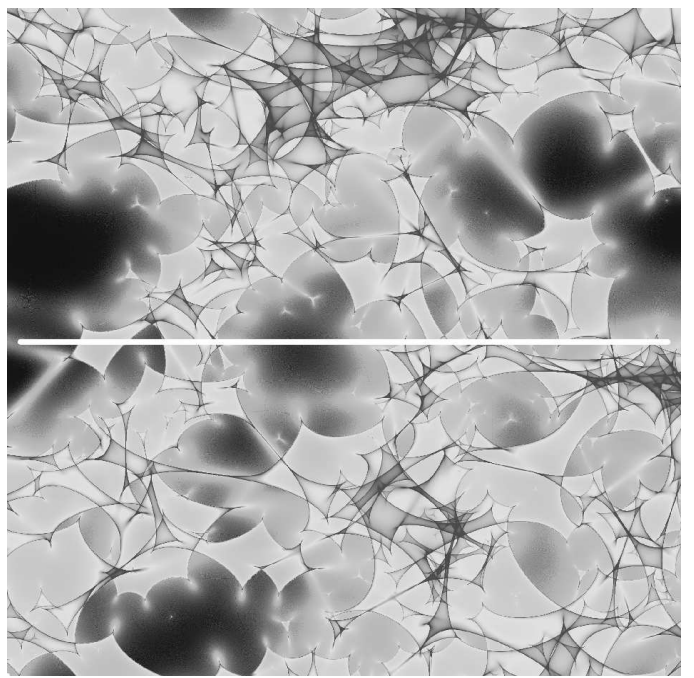}
\caption{Magnification map of Cloverleaf quasar H1413+117 image $A$.
A white solid line represents the analyzed path of an accretion disk center.}
\label{fig3}
\end{figure}

\section{Combined influence of intrinsic absorption and
gravitational microlensing on the Fe K$\alpha$ emission of QSOs}

To obtain modeled line profiles, it is necessary to define a number
of parameters which describe the line emitting region in the disk,
such as constraints for its size, the disk inclination angle and the
angular momentum of the central BH. For that purpose we usually use
the results from studies of the observed Fe K$\alpha$ line profiles
in AGN, such as e.g. \citet{nandra97,nandra07}. The inner radius
$R_\mathrm{in}$ of the disk cannot be smaller than the radius of the
marginally stable orbit $R_\mathrm{ms}$, that corresponds to
$R_\mathrm{ms}=6\ R_\mathrm{g}$ (gravitational radius
$R_\mathrm{g}=GM/c^2$, where $G$ is the gravitational constant, $M$
is the mass of central BH, and $c$ is the speed of light) in
Schwarzschild metric and to $R_\mathrm{ms}=1.23\ R_\mathrm{g}$ in
the case of Kerr metric with angular momentum parameter $a=0.998$.
To select the outer radius $R_\mathrm{out}$ of the disk, we take
into account some recent investigations of the Fe K$\alpha$ line
profile showing that it should be emitted from the innermost part of
the disk which outer radius is within several tens of R$_\mathrm{g}$
\citep[see e.g.][and references therein]{lcp06}. Therefore, in our
numerical simulations we adopted the following disk parameters:
inclination $i=35^o$, inner and outer radii $R_{in}=R_{ms}$ and
$R_{out}=20\ R_g$, emissivity index $q=-2.5$ and angular momentum of
central BH $a=0$ (Schwarzschild metric).

In order to study the influence of intrinsic absorption of
Cloverleaf quasar H1413+117 on its Fe K$\alpha$ line profile, we
modeled an X-ray absorption region using the following parameters:
$r_A = 0.2\ R_g << R_A = 7\ R_g$ (and thus $I_A(X,Y)$ can be assumed
as constant), central energy of absorption $E_A=E_0=6.4$ keV and
velocity dispersion $\sigma_E=0.5$ keV (that corresponds to random
velocity of clouds of $\sim 20000$ km/s). We assumed that the
absorption region is located at the following positions: (i)
$X_A=-15\ R_g$, $Y_A=0\ R_g$, (ii) $X_A=0\ R_g$, $Y_A=0\ R_g$ and
(iii) $X_A=15\ R_g$, $Y_A=0\ R_g$ and that it consists of 10000
individual absorbing clouds.

Microlensing magnification pattern for the image $A$ of H1413+117
(see Fig. \ref{fig3}) with 16 Einstein Ring Radii ($R_E$) on a side
(where $R_E\approx 1431\ R_g$) is calculated using the following
parameters: red shift of the lens: $z_d = 1.7$ \citep{chart04},
convergence $\kappa = 0.52$, shear $\gamma = 0.14$ \citep{kayser90}
and the masses of individual deflectors (stars) $m_d = 1\ M_\odot$.
Microlensing time scales, which are necessary for obtaining the
simulated light curves, are estimated from corresponding distance
scales according to the formula (13) of \citet{pj08}, in which
$R_{\rm source}$ is replaced by the distance from the center of
accretion disk, assuming a flat cosmological model with $\Omega
=0.3$ and $H_{0}= 75\ \rm km\ s^{-1} Mpc^{-1}$.

\section{Results}

\begin{figure*}
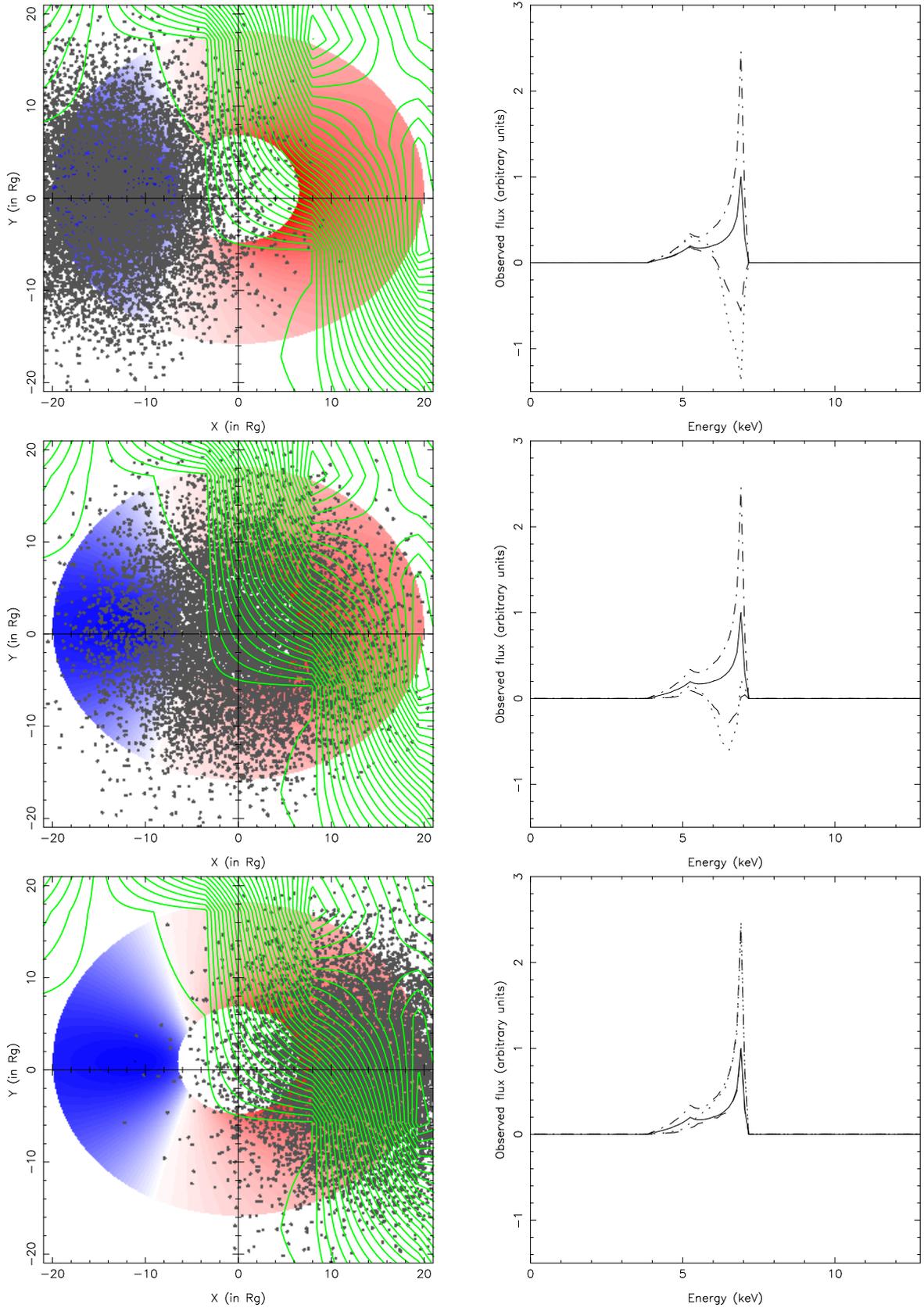

\centering
\includegraphics[width=0.84\textwidth]{fig4a.eps} \\
\includegraphics[width=0.84\textwidth]{fig4b.eps} \\
\includegraphics[width=0.84\textwidth]{fig4c.eps}
\caption{\emph{Left:} Relativistic accretion disk partially covered
by an X-ray absorption region (randomly scattered gray dots).
Magnification due to gravitation microlensing is also presented by
contour lines. Positions of absorption region are: $X_A=-15\ R_g$,
$Y_A=0\ R_g$ (top), $X_A=0\ R_g$, $Y_A=0\ R_g$ (middle) and $X_A=15\
R_g$, $Y_A=0\ R_g$ (bottom). \emph{Right:} Comparison between the
following corresponding Fe K$\alpha$ spectral line profiles:
unabsorbed (solid line), absorbed (dashed line), magnified
unabsorbed (dash-dot line) and  absorbed and magnified profile
(dotted line).} \label{fig4}
\end{figure*}

The results of numerical simulations of a relativistic accretion
disk partially covered by an X-ray absorption region where the Fe
K$\alpha$ line radiation is magnified due to gravitation
microlensing are presented in Fig. \ref{fig4} (left) for three
different positions of absorption region with respect to the
accretion disk center. In all three panels of Fig. \ref{fig4}, this
system consisting of accretion disk and absorption region is located
at the same position on the microlensing map of H1413+117 image $A$.
The corresponding comparisons between the unabsorbed, absorbed,
magnified and combined absorbed-magnified Fe K$\alpha$ spectral line
profiles are given in Fig. \ref{fig4} (right). As one can see from
Fig. \ref{fig4}, when X-ray radiation from approaching side of the
disk is significantly absorbed  there is a very strong absorption
component of the iron line and in such case the emission Fe
K$\alpha$ component looks redshifted at $\sim 5$ keV and is followed
by a strong absorption component at $\sim 7$ keV, indicating the P
Cygni profile of the iron line (see also Fig. \ref{fig2}).

Gravitational microlensing affects both components, but the
magnification of absorption component is much stronger (see dotted
line on the top panel of Fig. \ref{fig4}).

When the absorption region is located over the central part of the
disk, it also causes the occurrence of P Cygni profile of the iron
line, but in this case the intensity of absorption component is much
smaller, even when we take into account magnification due to
microlensing (see the middle panel of Fig. \ref{fig4}).

Finally, in the third case, when the absorption region is located
over the receding side of the disk, the absorption component of the
iron line is missing and the line profile is just slightly changed
due to intrinsic absorption. In the combination with gravitational
microlensing influence, it causes significant deformations of the Fe
K$\alpha$ line profile, but also without any indication for the
absorption component of the iron line (see the bottom panel of Fig.
\ref{fig4}).

\begin{figure*}
\centering
\includegraphics[width=0.8\textwidth]{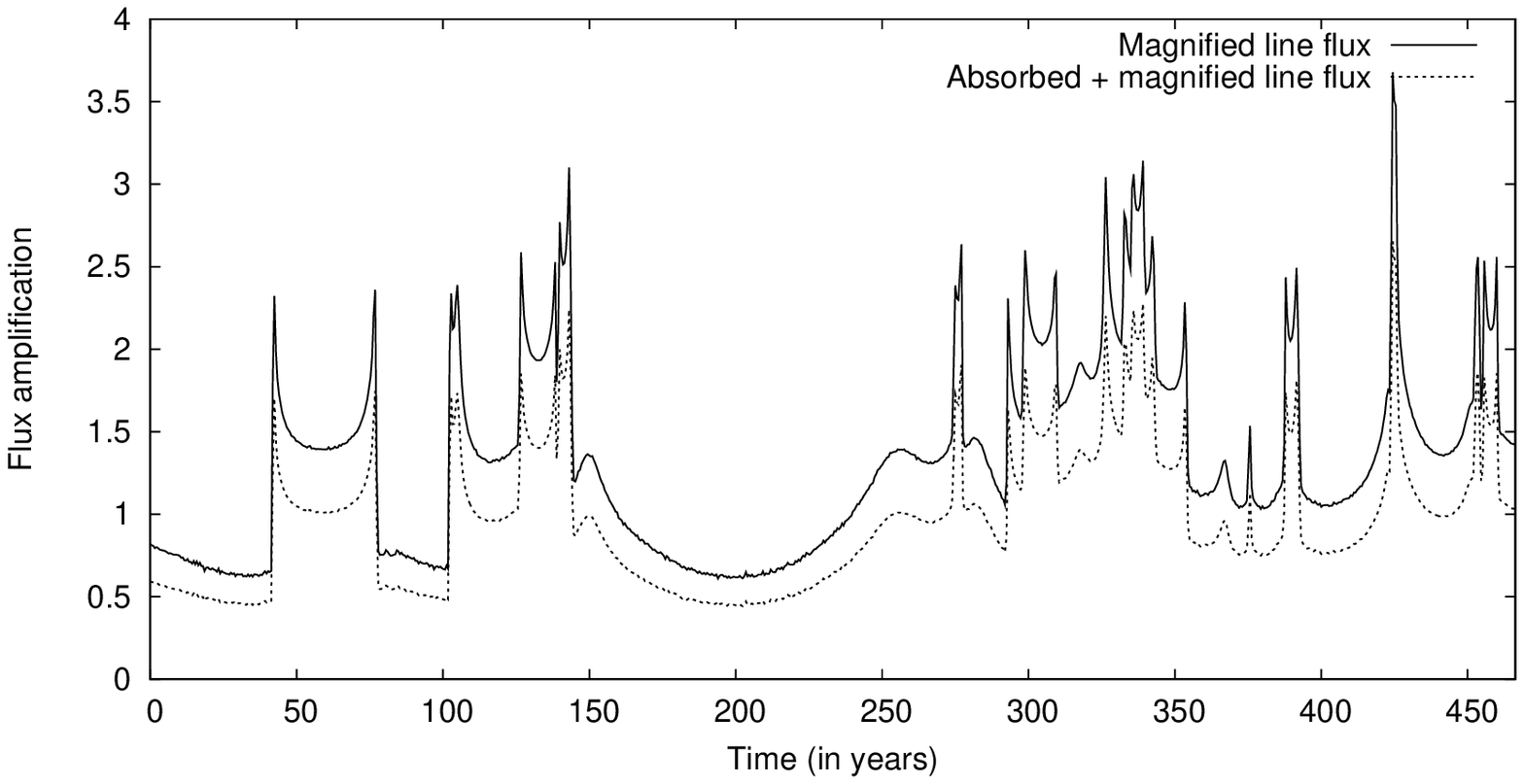} \\
\includegraphics[width=0.8\textwidth]{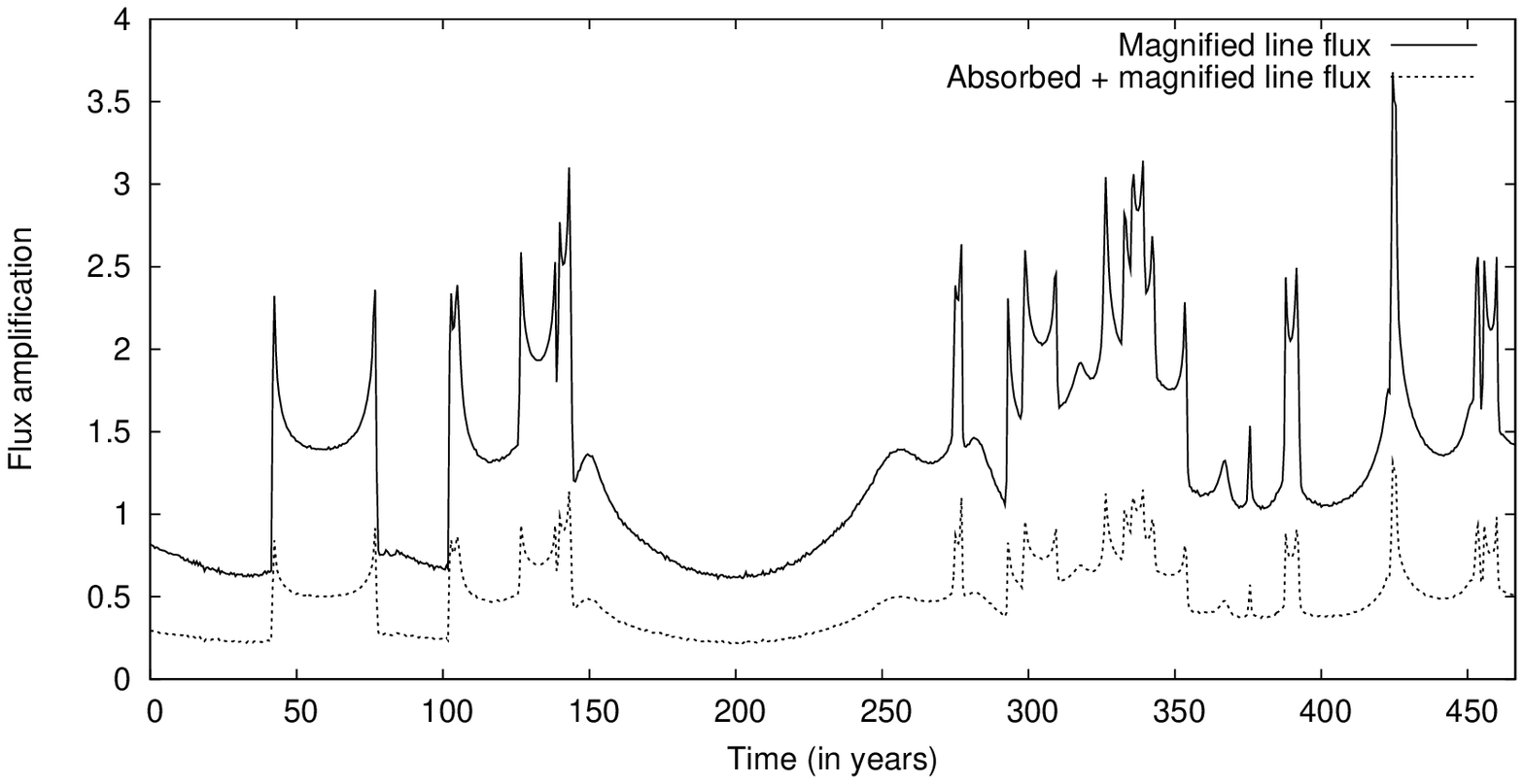} \\
\includegraphics[width=0.8\textwidth]{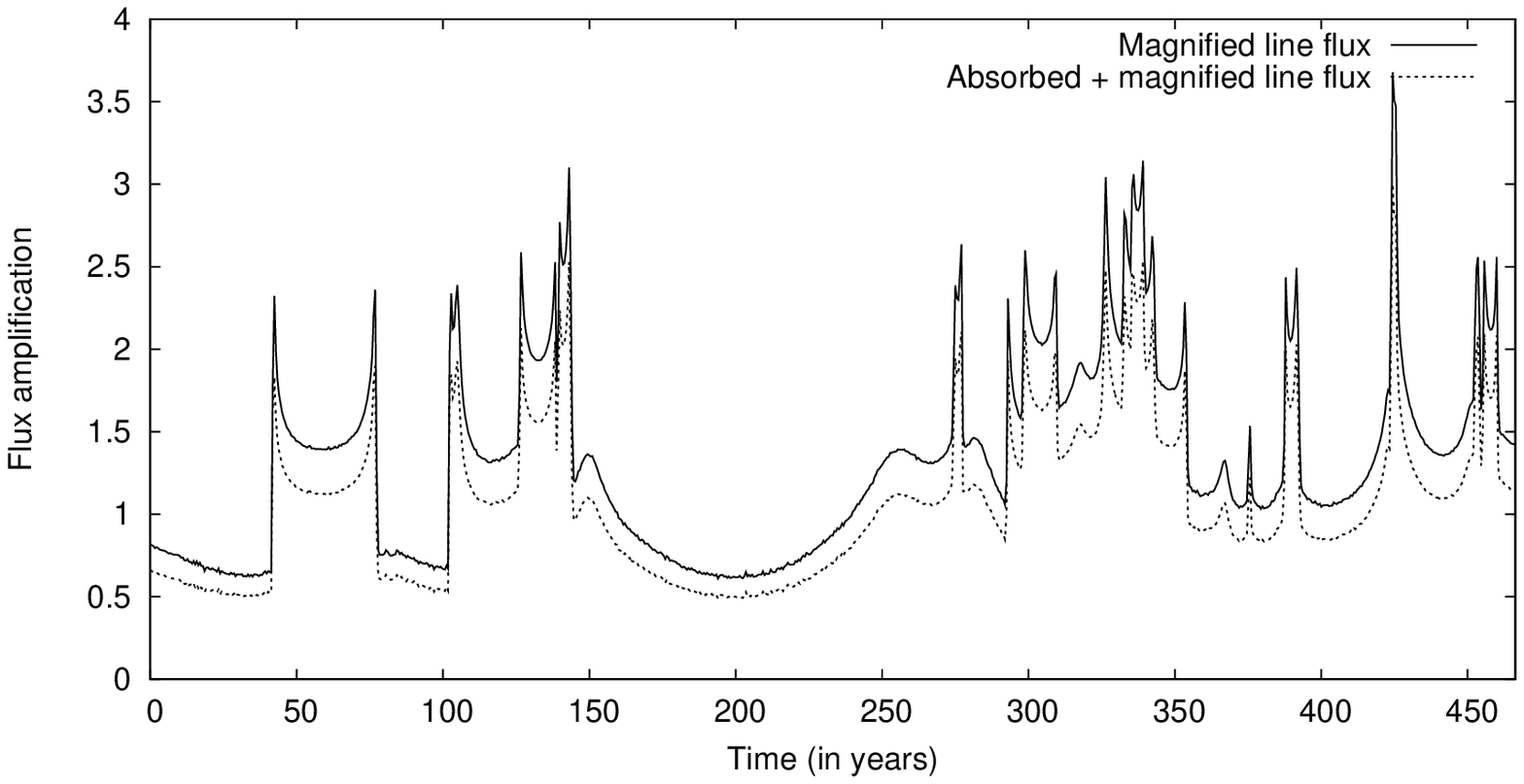}
\caption{Comparison between simulated light curves of unabsorbed
(solid line) and absorbed Fe K$\alpha$ line flux (dotted line).
Simulations are performed for the accretion disk crossing along the
horizontal path in the magnification map of image $A$ of Cloverleaf
quasar H1413+117 (see Fig. \ref{fig3}). The positions of absorption
region with respect to the accretion disk center are taken as:
$X_A=-15\ R_g$, $Y_A=0\ R_g$ (top), $X_A=0\ R_g$, $Y_A=0\ R_g$
(middle) and $X_A=15\ R_g$, $Y_A=0\ R_g$ (bottom).} \label{fig5}
\end{figure*}

The corresponding simulated light curves, produced when the system
consisting of the accretion disk absorbed emission  crosses over a
magnification pattern along the horizontal path in Fig. \ref{fig3},
are given in Fig. \ref{fig5}.

The three panels in Fig. \ref{fig5} show the simulated light curves
for three different positions of absorption region across the
accretion disk. We compared simulated light curves in the case when
the pure Fe K$\alpha$ emission is magnified (solid line) and when
absorbed emission of the Fe K$\alpha$ line is magnified (dashed
line). As it can be seen from Fig. \ref{fig5}  there is a global
correlation between these two light curves. The greatest difference
between the two light curves is in the case when absorption region
covers the central part of the disk and the smallest one is when it
covers the receding side of the disk. Such behavior of the light
curves is a consequence of different absorption rate of the iron
line for three different positions of absorption region.

From Fig. \ref{fig5} we can also estimate the frequency of so-called
high magnification events (HMEs - asymmetric peaks in the light
curves which depend not only on microlens parameters, but also on
parameters of the X-ray emitting region), i.e. the number of such
events per unit time. According to the obtained results, in the case
of image $A$ of H1413+117 one can expect one such event
approximately every 15 years, along the specified path. But, there
are also areas in the microlensing map of H1413+117A with a
significantly higher number of caustics (see e.g. top part of Fig.
\ref{fig3}), and therefore a source transit over these regions would
produce HMEs with much higher frequency. In principle, it can be
expected that the majority of HMEs should be detected in X-ray light
curves, less of them in UV and the smallest number in optical light
curves \citep{pj08}. \citet{pj08} found that the typical
microlensing time scales for the X-ray band are on order of several
months, while for the UV/optical bands they are on order of several
years.

\section{Conclusions}

In order to study how much the combined influence of absorption and
gravitational microlensing could change the Fe K$\alpha$ spectral
line profile, emitted from the X-ray emitting region of LoBAL QSOs
we developed a simple model of the X-ray absorbing/obscuring region
that could explain the observed P Cygni profile of the Fe K$\alpha$
line recently observed by \citet{done07}. We found that the model
can reproduce P Cygni profile only if at least a part of approaching
side of the disk is covered by absorbing material (see Fig.
\ref{fig2}). In particular case of 1H 0707-495, the observed sharp
drop at $\sim 7$ keV and emission at 5 keV can be explained by
absorption from relativistic, partially ionized material close to
the black hole that meanly cover the approaching side of the
relativistic accretion disk.

It is interesting that the width and depth of the absorption
component strongly depends on the projection of this region on the
accretion disk.

On the other side, this model (absorbed emission of the accretion
disk) has been used for simulation of gravitational microlensing of
such system by a random star field in the lens plane. We estimated
the magnification of the line flux and calculated the simulated
light curves produced when such source crosses over the H1413+117A
microlensing magnification map. Our results show that combined
influence of absorption and gravitational microlensing could explain
the observed variations of the Fe K$\alpha$ line profile in the case
of H1413+117 (emission features redward of the line rest-frame
energy and broad absorption features blueward of this energy)
reported by \citet{chart04}.

Note here, that we made a net of models for variations of the Fe
K$\alpha$ absorption of H1413+117A due to microlensing which can be
used for comparison with observed variations. We hope that these
results (and model) will be used in order to probe both, absorption
region and accretion disk characteristics in lensed QSOs with Fe
K$\alpha$ absorption component.

\bigskip\noindent\textit{Acknowledgements.} This work is a part of
the project (146002) "Astrophysical Spectroscopy of Extragalactic
Objects" supported by the Ministry of Science of Serbia.

\end{document}